\documentclass{pasj00}

\begin{document}
\SetRunningHead{M. Sasada et~al.}{Multiband photopolarimetric monitoring
of 3C~454.3}

\Received{2009/12/23}
\Accepted{2010/03/09}

\title{Multiband Photopolarimetric Monitoring of the Outburst of the
Blazar 3C~454.3 in 2007}

\author{
Mahito \textsc{Sasada}\altaffilmark{1},
Makoto \textsc{Uemura}\altaffilmark{2},
Akira \textsc{Arai}\altaffilmark{3},
Yasushi \textsc{Fukazawa}\altaffilmark{1},
Koji S. \textsc{Kawabata}\altaffilmark{2},\\
Takashi \textsc{Ohsugi}\altaffilmark{2},
Takuya \textsc{Yamashita}\altaffilmark{4},
Mizuki \textsc{Isogai}\altaffilmark{3},
Osamu \textsc{Nagae}\altaffilmark{1},
Takeshi \textsc{Uehara}\altaffilmark{1},\\
Tsunefumi \textsc{Mizuno}\altaffilmark{1},
Hideaki \textsc{Katagiri}\altaffilmark{1},
Hiromitsu \textsc{Takahashi}\altaffilmark{1},
Shuji \textsc{Sato}\altaffilmark{5},
and Masaru \textsc{Kino}\altaffilmark{5}}

\altaffiltext{1}{Department of Physical Science, Hiroshima University, Kagamiyama 1-3-1, Higashi-Hiroshima 739-8526}
\email{sasada@hep01.hepl.hiroshima-u.ac.jp}
\altaffiltext{2}{Astrophysical Science Center, Hiroshima
University, Kagamiyama 1-3-1, Higashi-Hiroshima 739-8526}
\altaffiltext{3}{Faculty of Science, Kyoto Sangyo University, Motoyama, Kamigamo, Kita-Ku, Kyoto-City 603-8555}
\altaffiltext{4}{National Astronomical Observatory of Japan, Osawa,
Mitaka, Tokyo 181-8588}
\altaffiltext{5}{Department of Physics, Nagoya University, Furo-cho, Chikusa-ku, Nagoya 464-8602}


%

\maketitle

\begin{abstract}
 We report on optical---near-infrared photopolarimetric observations of
 a blazar 3C~454.3 over 200~d. The object experienced an optical
 outburst in July 2007. This outburst was followed by a short state
 fainter than $V=15.2$~mag lasting $\sim 25$~d. The object, then, entered an
 active state during which we observed short flares having a timescale
 of 3--10~d. The object showed two types of features in the
 color--magnitude relationship. One is a ``bluer-when-brighter'' trend
 in the outburst state, and the other is a ``redder-when-brighter''
 trend in the faint state. These two types of features suggest a
 contribution of a thermal emission to the observed flux, as suspected
 in previous studies. Our polarimetric observation detected two
 episodes of the rotation of the polarization vector. The first one was
 a counterclockwise rotation in the $QU$ plane during the outburst
 state. After this rotation event of the polarization vector, the object
 entered a rapidly fading stage. The second one was seen in a series of
 flares during the active state. Each flare had a specific position
 angle of polarization, and it apparently rotated clockwise from the first
 to the last flares. Thus, the object exhibited rotations of the
 polarization vector in opposite directions. We estimated a decay
 timescale of the short flares during the active state, and then
 calculated an upper limit of the strength of the magnetic field,
 $B$=0.2~G, assuming a typical beaming factor of blazars,
 $\delta=20$. This upper limit of $B$ is smaller than those previously
 estimated from spectral analysis.
\end{abstract}

\section{Introduction}
Blazars are a subclass of active galactic nuclei (AGN), in which a
relativistic jet is viewed at a small angle to the line of sight. They
comprise two groups of objects, BL~Lac objects and Flat
Spectrum Radio-loud Quasars (FSRQs). Emission lines which originates
from the nuclear region (broad and narrow line regions) are observable
in the optical range in FSRQs, and rarely in BL~Lac objects
\citep{Urry95}.

The radio and optical fluxes from blazars are highly polarized because
synchrotron radiation from jets is dominant in the radio---optical
bands. Since the polarization can be a probe of the magnetic field in
the jet, observations of temporal variations of polarization are
important for the investigation of the structure of the jet. 
Rotations of polarization vectors have, in particular, attracted
attentions (e.g. \cite{Qian91}; \cite{Sillanpaa93}). \citet{Jones85}
proposed that an apparent rotation is a result of random motion of the
polarization vector. \citet{Marscher08} have recently reported a smooth
rotation of the polarization vector associated with an optical
flare in BL~Lac. They proposed that the rotation cannot be explained by
the random motion scenario, and indicates an emitting zone passing
through a helical magnetic field. The timescale of the rotation event in
BL~Lac is so short ($\sim 5$~d) that a number of similar events might
have been missed before. \citet{Larionov08} reported that the 
polarization vector in 3C~279 had rotated smoothly over approximately
two months. And another rotation event in PKS~1510$-$089 was reported by
\citet{Marscher10}, lasting 50$\pm$10~d. Long-lasting and high
time-resolved polarimetric observations are required to determine
whether observed rotations indicate real structures of the magnetic
field in jets.

3C~454.3 is classified as a FSRQ. \citet{Raiteri07} and
\citet{Raiteri08} reported a UV excess over the synchrotron component
in the spectral energy distribution (SED). The UV excess suggests a
substantial contribution of the thermal emission from an accretion
disk. Until $\sim$2001 only a moderate variability in a range of
$R\;\sim\;15-17$~mag was observed in the optical regime
\citep{Villata06}. The object showed an unprecedentedly bright optical
outburst in May 2005. \citet{Fuhrmann06} reported that the object
reached a maximum about $R\sim12.0$~mag during this bright outburst. The
object again experienced a major outburst in July---August 2007, and
subsequently another short flares in November 2007---February 2008
\citep{Raiteri08}. The object reached a maximum of $R=12.58$~mag during
the outburst state in 2007. In the gamma-ray region, a flare was also
detected by the AGILE satellite together with the 2007 optical outburst
\citep{Vercellone08}.

In general, a blazar becomes bluer when it is brighter. This feature is
so-called the ``bluer-when-brighter'' trend (e.g. \cite{Racine70}). On
the other hand, a ``redder-when-brighter'' trend has been observed in
3C~454.3 (\cite{Miller81}; \cite{Villata06}; \cite{Raiteri07}). This
trend showed a ``saturation'' at $R\sim14$~mag, and the object
possibly turned into a bluer-when-brighter trend in the bright states
during the 2007 outburst \citep{Raiteri08}. The redder-when-brighter
trend in 3C~454.3 is probably caused by a substantial contribution of
the thermal emission from the disk. Simultaneous optical and
near-infrared (NIR) observations could detect the bluer-when-brighter
trend in the bright state more clearly than the observations only
within the optical regime.

We performed photopolarimetric observations of 3C~454.3 simultaneously
in the optical and NIR bands from July 2007 to February 2008. In this
paper, we report the light curve, color and polarization variations of
3C~454.3 in the optical and NIR regions. Our polarimetric observations
found intriguing rotations of the polarization vector. This paper is
arranged as follows: In section 2, we present the observation method and
analysis. In section 3, we report the result of the photometric and
polarimetric observations. In section 4, first, we discuss the origin of
a long term component under the short flares. Second, we discuss
implications of the observed rotation events of the polarization vector,
Finally, we estimate the strength of magnetic field using a decay
timescale of flares. The conclusion is drawn in section~5.

\section{Observation}
We performed monitoring of 3C~454.3 from July 18, 2007 to February 2,
2008 using TRISPEC attached to the Kanata 1.5-m telescope at
Higashi-Hiroshima Observatory. TRISPEC (Triple Range Imager and
SPECtrograph) has a CCD and two InSb arrays, enabling photopolarimetric
observations in an optical and two NIR bands simultaneously
\citep{Watanabe05}. We used the photopolarimetric mode of TRISPEC with
$V$-, $R_{\rm C}$-, $I_{\rm C}$-, $J$-, and $K_{\rm S}$-band
filters. Although we obtained the polarimetric observation data in all
bands, we report the polarization parameter only in the $V$ band in this
paper. The data in the $V$, $J$ and $K_{\rm S}$ bands were observed in
the entire period. The data in the $R_{\rm C}$ and $I_{\rm C}$ bands were
obtained from JD~2454428 to 2454440. A unit of the observing sequence
consisted of successive exposures at four position angles of a half-wave
plate; 0\arcdeg, 45\arcdeg, 22.\arcdeg 5, 67.\arcdeg 5. A set of
polarization parameters was derived from each set of the four
exposures.

Integration time in each exposure varied night by night, depending
on the sky condition and the brightness of 3C~454.3. Typical
integration times were 90, 108, 108, 85 and 70~s in the $V$, $R_{\rm C}$,
$I_{\rm C}$, $J$ and $K_{\rm S}$ bands, respectively.

All images were bias-subtracted and flat-fielded, and we
performed aperture photometry with the {\it IRAF APPHOT} package. We
performed differential photometry with a comparison star taken in the
same frame of 3C~454.3. Its position is R.A.=$\timeform{22h53m58s.11}$,
Dec.=$\timeform{+16D09'07''.0}$ (J2000.0) and its magnitudes are 
$V = 13.587$, $R_{\rm C} = 13.035$, $I_{\rm C} = 12.545$, $J = 11.858$
and $K_{\rm S} = 11.241$~mag (\cite{Gonzalez-perez01};
\cite{Skrutskie06}). The comparison star is listed in \citet{Fiorucci98}
as star H. The position of a check star is 
R.A.=$\timeform{22h53m44s.63}$, Dec.=$\timeform{+16D09'08''.1}$ listed
in \citet{Gonzalez-perez01} as star 14. Using the check star, we
confirmed that the comparison star was almost constant in magnitude
within 0.05~mag during our observation period. We calculated the
magnitude of 3C~454.3 using another neighboring stars listed in
\citet{Gonzalez-perez01}, and estimated a systematic error of the
magnitude of 3C~454.3 depending on the comparison stars to be 
$\sim 0.1$~mag in all photometric bands. In this paper, the error of the
flux of 3C~454.3 in figure~6 includes both the systematic and photon
statistical errors, and that in the other figures includes only the
photon statistical error.

We confirmed that the instrumental polarization was smaller than 0.1 \%
in the $V$ band using the observation of unpolarized standard stars. We,
hence, applied no correction for it. The zero point of the polarization
angle is corrected as the standard system (measured from north to east)
by observing the polarized stars, HD~19820 and HD~25443
\citep{Wolff96}.

\section{Result}
\subsection{Optical and NIR light curves}
\begin{figure*}
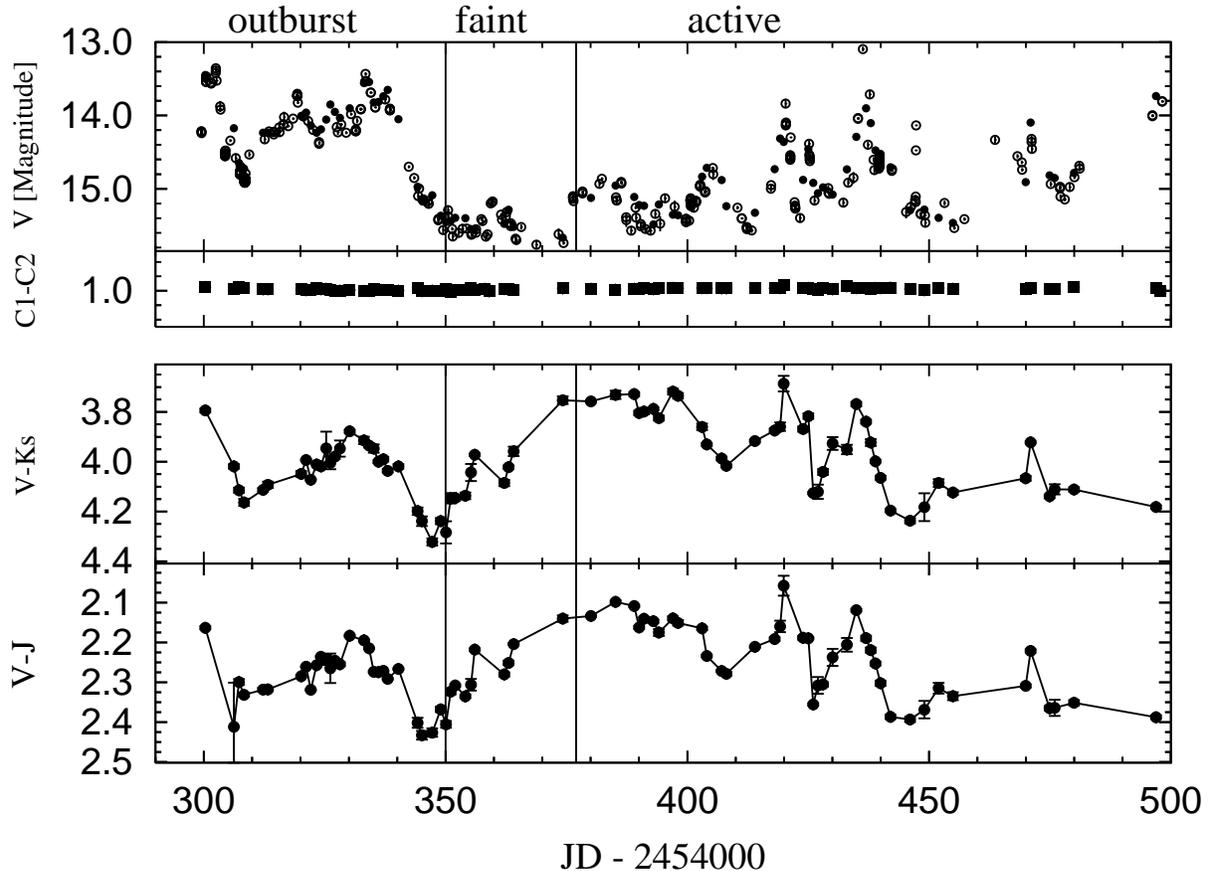

  \begin{center}
 \FigureFile(180mm,75mm){fig1.eps}
  \end{center}
  \caption{%
  Light curve of 3C~454.3 in the $V$ band, color variations of 
 $V-K_{\rm S}$ and $V-J$. The filled and open circle denote the
 observation with the Kanata telescope and by the WEBT team,
 respectively. We also show the relative magnitude, C1-C2,
 between the comparison star (C1) and the check star (C2). The three
 states, that is, ``outburst'', ``faint'' and ``active'' states, can be
 defined as indicated by the vertical solid lines (for detail, see the
 text).}%
  \label{fig1}
\end{figure*}

Figure~1 shows the light curve in the $V$ band, the color variations in
$V-K_{\rm S}$ and $V-J$. We also show the relative magnitude, C1-C2,
between the comparison star (C1) and the check star (C2). We can
recognize three states of the object based on the temporal behavior; the
first is an outburst state from JD~2454300 to 2454350, the second is a
faint state from JD~2454350 to 2454374, and the last is an active state
from JD~2454374 to 2454500.

According to \citet{Raiteri08}, the outburst started on $\sim$
JD~2454270, and reached a maximum of $R\sim 12.7$~mag in JD~2454300. We
found $V\sim 13.2$~mag on the same night (when we started our
monitoring). After the maximum, we detected a short fading, which was
followed by a rebrightening trend. Several short flares were
superimposed on the rebrightening trend, for example on JD~2454326 and
2454334. The object exhibited a rapid fading from the outburst state
from JD~2454345, and then became faint to $V=15.2$~mag within 10~d.

In May 2005, the object experienced an exceptionally large outburst
\citep{Fuhrmann06}. The object reached a maximum about $V=12.7$ and
$R=12.2$~mag. In the case of the 2005 outburst, it took $\sim 75$~d to
decay from the outburst maximum to the faint state of $R=15.8$~mag
\citep{Villata06}. The 2005 outburst, thus, faded much more slowly than
the 2007 outburst.

After the rapid fading of the 2007 outburst, the object entered the
faint state lasting 25~d. The criterion between the outburst and
faint states is also based on different color--magnitude relationships,
as reported in subsection 3.2. The object was relatively inactive in the
$V$~band, showing no significant variation over 0.6~mag during the
faint state.

\subsection{Color Variation}
\begin{figure}
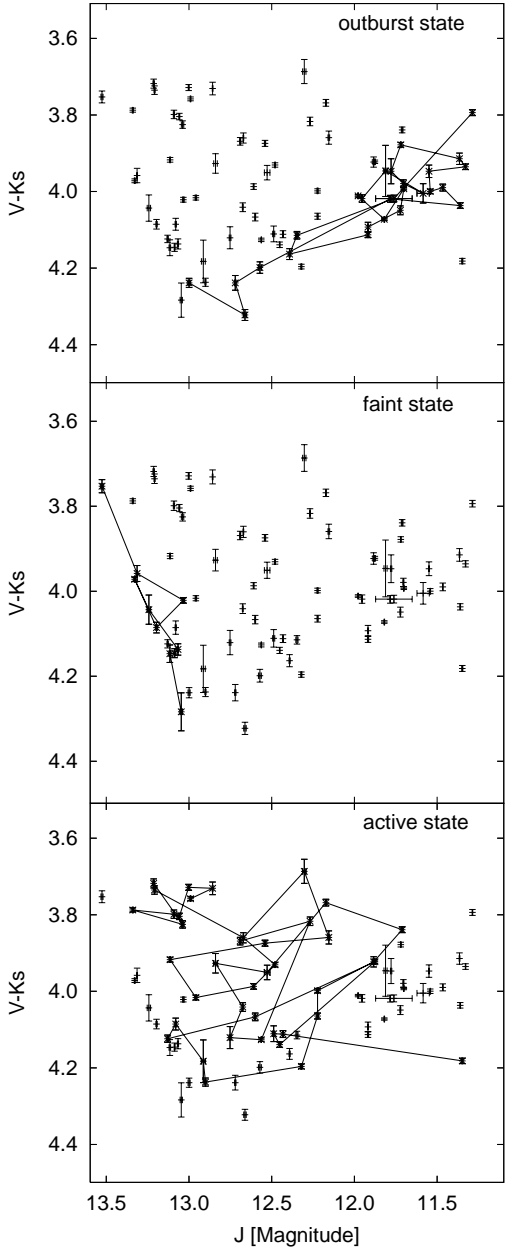

  \begin{center}
 \FigureFile(80mm,100mm){fig2.eps}
  \end{center}
  \caption{%
 Color--magnitude diagram in the $J$-band magnitude and
 $V-K_{\rm S}$ color. The outburst, faint and active states are
 indicated by the solid lines connecting the observed points in the top,
 middle and bottom panels, respectively.}%
  \label{fig2}
\end{figure}

Figure~2 shows the color--magnitude relationships in the outburst (top
panel), the faint (middle panel) and the active states (bottom
panel). In the outburst state, the color well correlated with the
$V$-band flux both in the long outburst trend and the short flares
superimposed on the outburst. The object exhibited a
bluer-when-brighter trend. A rapid reddening was associated with the
rapid fading from the outburst. At the beginning of the faint state, the
object was reddest during our observation period. 

The color--magnitude relationship in the faint state is totally
different from that in the outburst state, as clearly shown in
figure~2. The $V$-band flux kept a slow decline in the faint state,
although the color showed a rapidly bluing trend. In contrast to the
outburst state, the faint state showed a redder-when-brighter
trend. 

In the active state, a bluer-when-brighter trend was associated with
the short flares. The object was quite bright both on JD~2454300 and
2454498. The color on JD~2454498 was, however, significantly redder than
that on JD~2454300, as can be seen in figure~1. This color behavior
suggests the presence of a long-term reddening trend during the active
state. The color--magnitude relation in the active state is apparently
complex compared with those in the outburst and faint states, as shown
in figure~2. This can be interpreted as a result of the superposition of
two components, namely the short flares with a bluer-when-brighter
trend and the long term component with a reddening trend.

\citet{Raiteri07} reported the presence of a thermal component in the
SED of 3C~454.3. The redder-when-brighter trend of 3C~454.3 in the
faint state is, hence, interpreted as the increase of the contribution
of the synchrotron emission compared with the thermal emission, when the
flux increased. The reddening feature of the long term component in the
active state suggests that the contribution of the synchrotron emission
increased with time.

\citet{Raiteri08} reported on the $B-R$ color variation associated with
the flux variation between 2005---2008. In their observation, 3C~454.3
generally exhibited a redder-when-brighter trend, while the trend
weakened when the object was brighter than $R \sim 14$~mag. The color
behavior obtained by our observations is generally consistent with that
reported in \citet{Raiteri08}, except for the clear bluer-when-brighter
trend detected in our observation. We propose two reasons why the
bluer-when-brighter trend was prominent in our observations. One
reason is that the contribution of the synchrotron emission is larger in
the NIR region than in the optical. The other is that the range of
wavelength between $V$ and $J$ bands are larger than that between $B$
and $R$ bands. An observation with a wider range of the wavelength would
detect a small color-variation more readily.

\subsection{Polarization parameters}
\begin{figure}
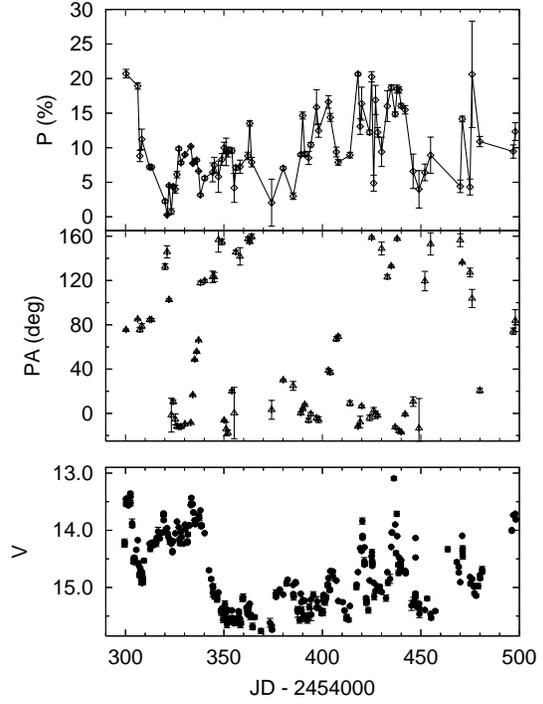

  \begin{center}
 \FigureFile(80mm,65mm){fig3.eps}
  \end{center}
  \caption{%
  Temporal variation of the polarization of 3C~454.3; (Top) polarization
 degree (\%) in the $V$~band, (middle) polarization angle (degree) in
 the $V$~band, and (bottom) $V$-band light curve. We set the range of
 the polarization angle from -20$^{\circ}$ to 160$^{\circ}$. }%
  \label{fig3}
\end{figure}

Figure~3 shows the temporal variation of the polarization degree (top),
the polarization angle (middle) and the light curve (bottom) in the
$V$~band. Figure~4 shows temporal variations of the polarization vector
in the $QU$ planes. The observed flux was corrected for the interstellar 
absorption in our Galaxy in the middle and bottom panels of 
figure~4 \citep{Schlegel98}. The $V$-band light curve is shown in the
top panel of figure~4 just for reference. In the middle and bottom
panels, the points during the outburst and the active states are
connected with the solid and dashed lines, respectively. We calculated
Stokes parameters $Q$ and $U$ using the energy flux, $I$
($={\nu}F_{\nu}$), and $Q/I$ and $U/I$ obtained from the observation.

The polarization degree was quite high ($20.7\pm0.6$~\%) in the
early outburst state, then rapidly decreased to almost 0~\% within
21~d. The polarization vector subsequently showed a smooth
counterclockwise rotation in the $QU$ plane, as shown in the middle
panel of figure~4. This rotation episode lasted for 19~d from
JD~2454328. The rotation speed was fairly constant with 
$d{\rm PA}/d{\rm t}=22\pm3$ degree/d particularly from JD~2454333 to
2454338. It is evidently difficult to explain this systematic rotation
by a result of a random walk in the $QU$ plane \citep{Jones85}. The
object entered the rapid fading stage from the outburst after that
rotation of the polarization vector. 

Another rotation episode of the polarization vector was observed during
the active state. We labeled six flares in the active state from ``A''
to ``F'', as indicated in the top and bottom panels of figure~4. We
calculated the averages of $Q$ and $U$ during each flare, which is indicated by
the open diamonds in the bottom panel. The filled square in the bottom 
panel denotes the averages of $Q$ and $U$ of the bottom of the
flares. The averages of $Q$ and $U$ in each flare were significantly
apart from those of the bottom of the flares, except for flare~A. The
directions of polarization were different in flare by flare. Our
observation, thus, indicates that a specific polarization vector was
associated with those flares. Furthermore, the average polarization
angles of the flares apparently rotate in a clockwise direction from the
first to the last.

The first counterclockwise rotation in the outburst state is analogous
to the rotation which has recently been observed in BL~Lac in terms of the
timescale and the smoothness of the variation in the polarization angle
\citep{Marscher08}. The second clockwise rotation was an unprecedented
event in which a flare-by-flare rotation was observed. This is also the
first time that the two rotations of the polarization vector were
consecutively observed, and furthermore, in opposite directions.

As well as the increase of the polarized flux, the increase of the
polarization degree was likely to be associated with the short flares in
the active state, as can be seen in figure~3. On the other hand, no such
a trend can be seen for the long-term reddening component in the active
state. For example, the polarization degrees in $\sim$JD~2454500 were
comparable with those of the bottoms of the short flares in
JD~2454400--2454430, while the total flux was much higher in
$\sim$JD~2454500 than that in the short flare in
JD~2454400--2454430.

\begin{figure}
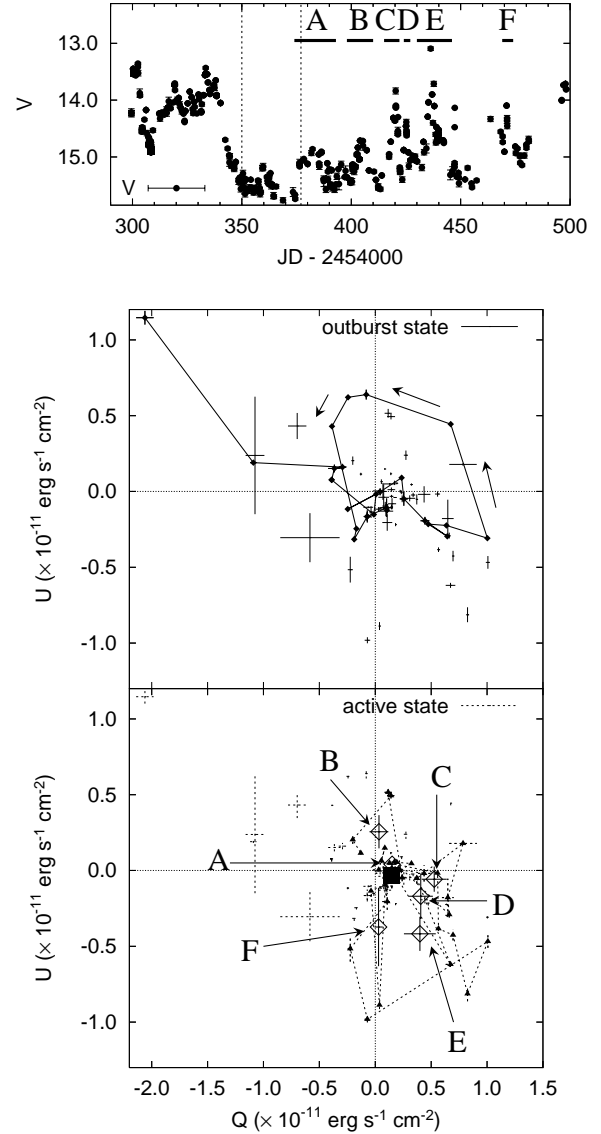

  \begin{center}
 \FigureFile(80mm,65mm){fig4.eps}
  \end{center}
  \caption{%
 Top panel: Light curve of the object in the $V$ band. The period of the
 outburst, faint, and active states were indicated by the vertical
 dashed lines. In the active state, we identified 6 flares, that is,
 flare A (from JD~2454374 to 2454393), B (from JD~2454398 to 2454408), C
 (from JD~2454413 to 2454423), D (from JD~2454424 to 2454427), E (from
 JD~2454430 to 2454446) and F (from JD~2454470 to 2454475). Middle
 panel: The $QU$ plane during the outburst state in the $V$ band. Bottom
 panel: The $QU$ plane during the active state in the same band. The
 open squares indicate the averaged $QU$ points during each flare. The
 flux was calculated with 1.98 
 $\times 10^{-5} {\rm erg\ sec^{-1}\ cm^{-2}}$ at 0~mag in $V$ band
 \citep{Fukugita95}}%
  \label{fig4}
\end{figure}

\subsection{Temporal variation of the spectral energy distribution
  associated with a short flare}
We performed $V$-, $R_{\rm C}$-, $I_{\rm C}$-, $J$-, $K_{\rm S}$-band
photometry during the short flare ``E''. In this section, we report the
temporal variation of the SED during this flare.

\begin{figure}
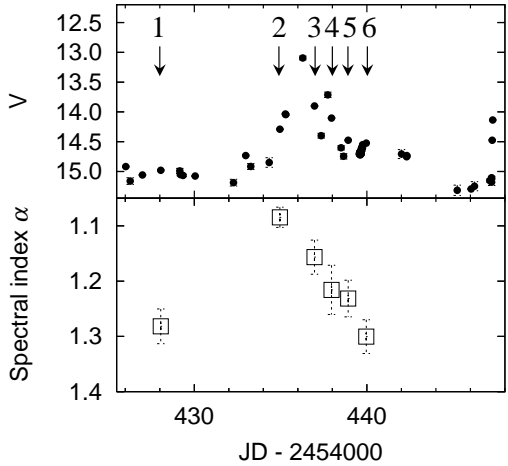

  \begin{center}
 \FigureFile(80mm,65mm){fig5.eps}
  \end{center}
  \caption{%
 Upper panel: Light curve of flare ''E'' in the $V$ band. The labels
 from 1 to 6 indicate the epochs in which we performed 5-band multicolor
 observations. Lower panel: Temporal variation of spectral index
 $\alpha$ from epoch~1 to 6.}%
  \label{fig5}
\end{figure}

\begin{figure}
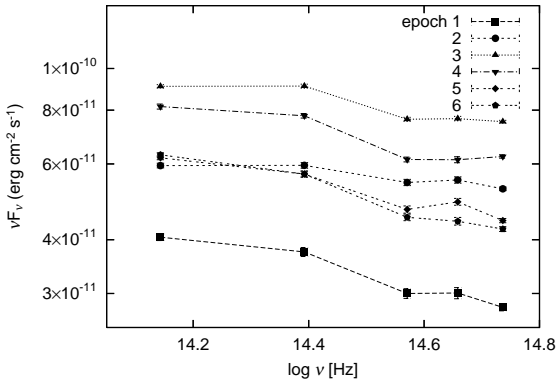

  \begin{center}
 \FigureFile(80mm,65mm){fig6.eps}
  \end{center}
  \caption{%
 SEDs from the $V$ to the $K_{\rm S}$ bands at each epoch. The vertical axis
 indicates the flux, and the unit is 
 ${\rm erg\ sec^{-1}\ cm^{-2}}$. These fluxes were calculated with 1.42,
 0.895, 0.384 and 0.0867 $\times 10^{-5} {\rm erg\ sec^{-1}\ cm^{-2}}$
 at 0~mag in $R_{\rm C}$, $I_{\rm C}$, $J$ and $K_{\rm S}$ bands,
 respectively (\cite{Fukugita95} and \cite{Bessell98}).}%
  \label{fig6}
\end{figure}

The upper and lower panels of figure~5 show the $V$-band light curve and
the temporal variation of the spectral index during flare ``E'',
respectively. Six epochs from 1 to 6 are indicated, in which the 5-band
photometric data are available. We calculated the spectral index,
$\alpha$, using the 5-band absorption-corrected data, assuming a
power-law form of SEDs, that is, $F_{\nu}\propto\nu^{-\alpha}$. In order
to ignore fine structures of the spectra, we fitted the data with errors
of 10~\% of the flux. Epoch~2 was in a rising phase of the
flare. Epoch~3, 4, 5 and 6 were in a decay phase.

As can be seen in figure~5 and 6, the epoch with the hardest SED
precedes the peak of the $V$-band flux. \citet{Takahashi96} reported the
X-ray spectral evolution associated with a flare of Mrk~421 whose
timescale is similar to that of flare ``E'' in 3C~454.3. According to
\citet{Takahashi96}, the X-ray spectrum was hardest before the flare
maximum, as observed in our optical---NIR observations of
3C~454.3. Mrk~421 is one of objects called as high frequency BL Lac
objects, in which the synchrotron peak frequency is in the X-ray
range. On the other hand, the synchrotron peak frequency of 3C~454.3 is
in the infrared range. The SED evolutions of the synchrotron flares are
remarkably analogous in Mrk~421 and 3C~454.3, although the peak
frequencies of the synchrotron radiation are totally different.

\section{Discussion}
\subsection{Long term component during the active state}
We found distinct two components during the active state, namely, the
short flares and the long-term reddening component. The flares in the
active state had short timescales, the bluer-when-brighter trend and
specific polarization components. Compared with these features of the
flares, the long term component had a long timescale, reddening trend
and no correlation of the polarization degree. 

The size of the emitting zone of the long term component would be
larger than that of the short flares if the observed timescale
represents the light travel time of the emitting zone. The observed
behavior of polarization may be interpreted with the difference in the
emitting region size. The magnetic field in various directions possibly
exists in a large emitting area. In this case, the polarization of a
certain area is diluted by those of another areas, and thereby, the
polarization degree may be insensitive to the flux variation, as
observed in the long term component. On the other hand, the direction of
the magnetic field is expected to be relatively ordered in a small, or
local area. Flares from such a small area would be observed with an
increase of the polarization degree, as the short flares in the active
state.

\subsection{Rotations of the polarization vector}
The object exhibited two episodes of polarization-vector rotations
both in the clockwise and counterclockwise directions in the $QU$
plane. \citet{Marscher08} proposed that the rotation of the
polarization vector occurs when the optical emitting zone propagates
outward along a helical magnetic field in the jet. As mentioned in
subsection~3.3, the rotation of the polarization vector in the outburst
state in 3C~454.3 is reminiscent of that in BL~Lac observed by
\citet{Marscher08}. In the case of the clockwise rotation in the active
state, the emitting zone should shift to inverse direction compared
with the counterclockwise rotation in the outburst, if the scenario
proposed in \citet{Marscher08} is applied to our result. In other words,
our results require both outward and inward shifts of the emitting
zone in the jet. However, it is difficult to consider an
upstream propagation of a shock or emitting matter in the jet. An
apparent inward shift of the emitting zone may be possible if the
location of the emitting zone systematically varies depending on the
physical condition in the jet. For example, if the dominant emitting
zone locates in the downstream of the jet during an outburst, the
emitting zone can apparently shift to the upstream as the outburst
finishes. If this is the case for 3C~454.3, similar rotations of the
polarization vector should be confirmed in every outbursts.

We can consider an alternative scenario that the polarization vector
actually rotated in the counterclockwise direction in the active state,
because the observed polarization angle has an ambiguity of 180
degree. In the outburst state, the rotation rate of the polarization angle
was 22 degree/d as mentioned subsection~3.2. Figure~7 shows the temporal
variation of the polarization angle of the short flares. Each point shows
the averaged polarization angle of each flare. In the case of the
clockwise rotation, the rotation rate is estimated to be 1.1$\pm$0.3
degree/d. The solid line in figure~7 indicates the best-fitted model of
the clockwise rotation with a constant rotation rate. The rotation rate
is much smaller than that in the outburst state. On the other hand, the
dashed line in figure~7 indicates the model of the counterclockwise
rotation. In this case, the rotation rate is estimated to be
12.1$\pm$0.9 degree/d, which is relatively close to the rate in the
outburst state. Hence, the two rotation episodes of the polarization
vector in 3C~454.3 could have the same origin under a helical magnetic
field, if the polarization angle actually rotated in the
counterclockwise direction during the active state. 

\begin{figure}
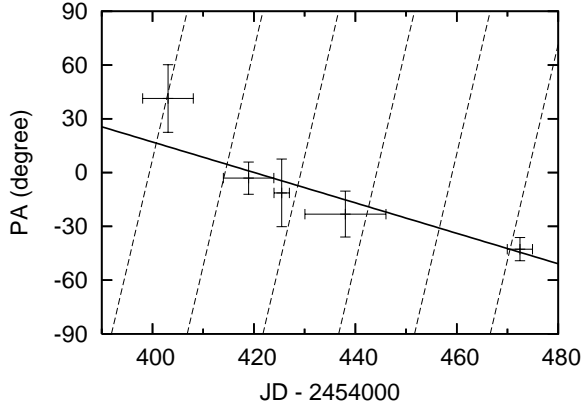

  \begin{center}
 \FigureFile(80mm,65mm){fig7.eps}
  \end{center}
  \caption{%
  Temporal variation of the polarization angle of the short flares
 during the active state. The solid line shows the best-fitted model of
 the clockwise rotation. The dashed line shows that of counterclockwise
 rotation.}%
  \label{fig7}
\end{figure}

\subsection{Estimation of the strength of the magnetic field in the
  emitting zone}
We can estimate the strength of the magnetic field in the emitting zone
using the decay timescale of observed flares. We estimate the decay
timescale from flares in the active state. In figure~8, we show the
light curves of flares whose timescales are of the shortest among our
observed flares. We identified five decay phases, as labeled ``a'',
``b'', ``c'', ``d'' and ``e'' in figure~8. We calculated the decay
timescale of the flares, $\tau_{\rm obs}$, assuming that the flux
followed an exponential decay, that is, 
$F(t) \propto e^{-t/\tau_{\rm obs}}$ \citep{Bottcher07}. The decay
timescales were calculated to be 1.9$\pm$0.2, 1.9$\pm$0.2,
0.89$\pm$0.02, 1.0$\pm$0.1 and 0.9$\pm$0.1~d for the decay phases ``a'',
``b'', ``c'', ``d'' and ``e'' in figure~8, respectively.

\begin{figure}
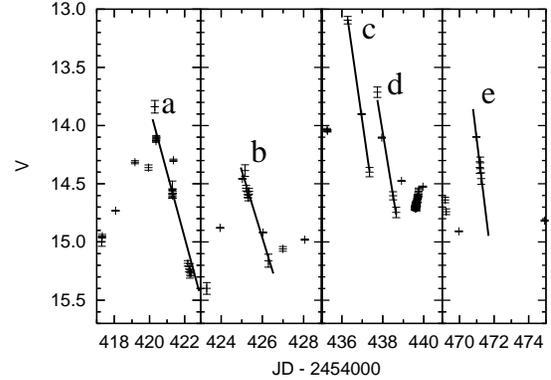

  \begin{center}
 \FigureFile(80mm,65mm){fig8.eps}
  \end{center}
  \caption{%
 $V$-band light curves of the short flares in the active state. The
 solid lines shows the best-fitted exponential decay model. }%
  \label{fig8}
\end{figure}

We can estimate the strength of a magnetic field, $B$, in the emitting
zone using $\tau_{\rm obs}$. \citet{Tashiro95} gave the synchrotron
cooling timescale, $\tau_{\rm s}$, for relativistic jets under
homogeneous magnetic field, as follows:
\begin{eqnarray}
\tau_{\rm s} \sim 3.2 \times 10^{4} \times B^{-3/2} E_{\rm ph}^{-1/2}
 \delta^{-1/2}\ [s],
\end{eqnarray}
where $B$ and $\delta$ are the strength of the magnetic field in
Gauss~(G) and the beaming factor $\delta$, respectively. $E_{\rm ph}$ is
the emitting photon energy in the observer's frame. It is 2.25~eV in the
case of our $V$-band observation. Following \citet{Tashiro95}, we
estimated $\tau_{\rm s}$ with the observed decay timescale in which the
cosmological effect is corrected, to be $\tau_{\rm obs}/(1+z)$. Using a
typical beaming factor for blazars, $\delta=20$, we calculated the
strength of the magnetic field $B$, to be 0.14, 0.14, 0.24, 0.22 and
0.24~G from the decay phase ``a'', ``b'', ``c'',''d'' and ``e'',
respectively \citep{Burbidge74}. Thus, the estimated strength of the
magnetic field using the decay timescale was typically 0.1--0.2~G.

This estimation actually provides an upper limit of $B$ because the
Compton cooling effect possibly makes $\tau_{\rm obs}$ smaller than a
real $\tau_{\rm s}$ (e.g. \cite{Bottcher07}). According to
\citet{Vercellone09}, the luminosity of the inverse-Compton scattering
component was 10 times larger than that of the synchrotron component in
3C~454.3. This implies that a significant Compton cooling actually
occurs. Hence, an accurate calculation of $B$ requires the estimation of
the Compton cooling effect through a spectral analysis including the
inverse-Compton component. Observations in $\gamma$-ray range, for
example, with Fermi Gamma-ray Space Telescope, are important to
investigate the inverse-Compton component, and thereby, estimate $B$.

The strength of the magnetic field in 3C~454.3 has been estimated to be
from 0.3 to 35~G based on the spectral analysis of multi-wavelength SEDs
(e.g. \cite{Ghisellini07}; \cite{Sikora08}). Our estimate of the upper
limit of $B$ is smaller than those from spectral analysis.
We can consider two reasons; one is that the beaming factor, $\delta$,
could be atypically smaller. If we assume an atypically small value of
$\delta=2$, $B$ is estimated to be 0.3--0.5~G, which is consistent with
typical value from spectral analysis. The other is that we may
overestimate the cooling timescale $\tau_{\rm obs}$ if the observed
flares was actually a superposition of several shorter flares, and then,
underestimate the upper limit of $B$.

\section{Conclusion}
We observed the outburst, faint and post-outburst active states of
the blazar 3C~454.3 in 2007 in multicolor photopolarimetric mode. We
found that 3C~454.3 experienced the rotations of the polarization vector
in both clockwise and counterclockwise directions in the $QU$ plane
during the outburst and post-outburst active states. The feature of the
bluer-when-brighter trend was observed during the outburst state. On the
other hand, the feature of the redder-when-brighter trend was observed
during the faint state. It indicates that the relative contribution of
the thermal component increased in the $V$ band during this state. 
\\
\\
This work was partly supported by a Grant-in-Aid from the Ministry of
Education, Culture, Sports, Science and Technology of Japan (19740104).
A part of the light curve presented in this paper is based on data taken
and assembled by the WEBT collaboration and stored in the WEBT archive
at the Osservatorio Astronomico di Torino-INAF
(http://www.oato.inaf.it/blazars/webt/).


\begin{thebibliography}{}
\bibitem[{Bessell,} {Castelli} {\&} {Plez}(1998)]{Bessell98}
  Bessell,~M.~S.,\ Castelli,~F.,\ \& Plez,~B.\ 1998, \aap, 333, 231 
\bibitem[{${\rm B\ddot{o}ttcher}$} {et~al.}(2007)]{Bottcher07}
  ${\rm B\ddot{o}ttcher}$,~M.\ \etal\ 2007, \apj, 670, 968 
\bibitem[{Burbidge,} {Jones} {\&} {Odell}(1974)]{Burbidge74}
  Burbidge,~G.~R.,\ Jones,~T.~W.,\ \& Odell,~S.~L.\ 1974, \apj, 193, 43 
\bibitem[{Fiorucci} {et~al.}(1998)]{Fiorucci98}
  Fiorucci,~M.,\ Tosti,~G.,\ \& Rizzi,~N.\ 1998, \pasp, 110, 105 
\bibitem[{Fuhrmann} {et~al.}(2006)]{Fuhrmann06}
  Fuhrmann,~L.,\ \etal\ 2006, \aap, 445, L1 
\bibitem[{Fukugita} {et~al.}(1995)]{Fukugita95}
  Fukugita,~M.,\ Shimasaku,~K.,\ \& Ichikawa,~T.\ 1995, \pasp, 107, 945
\bibitem[{Ghisellini} {et~al.}(2007)]{Ghisellini07}
  Ghisellini,~G.,\ Foschini,~L.,\ Tavecchio,~F., \& Pian,~E.\ 2007, \mnras, 382, L82
\bibitem[{${\rm Gonz\acute{a}lez-P\acute{e}rez}$}
			   {et~al.}(2001)]{Gonzalez-perez01}
			   ${\rm Gonz\acute{a}lez-P\acute{e}rez}$,~J.~N.,\
			   Kidger,~M.~R.,\ \& 
			   ${\rm Mart\acute{\i}n-Luis}$,~F.\ 
			   2001, \aj, 122, 2055
\bibitem[{Jones} {et~al.}(1985)]{Jones85}
  Jones,~T.~W.,\ Rudnick,~L.,\ Fiedler,~R.~L.,\ Aller,~H.~D.,\
			   Aller,~M.~F.,\ \& Hodge,~P.~E.\ 1985, \apj,
			   290, 627
\bibitem[{Larionov} {et~al.}(2008)]{Larionov08}
  Larionov,~V.~M.,\ \etal\ 2008,\ \aap, 492, L389 
\bibitem[{Marscher} {et~al.}(2008)]{Marscher08}
  Marscher,~A.~P.,\ \etal\ 2008,\ \nat, 452, 966 
\bibitem[{Marscher} {et~al.}(2010)]{Marscher10}
  Marscher,~A.~P.,\ \etal\ 2010,\ \apj, 710, L126 
\bibitem[{Miller} (1981)]{Miller81}
  Miller,~H.~R. 1981,\ \apj, 244, 426 
\bibitem[{Qian} {et~al.}(1991)]{Qian91}
  Qian,~S.~J.,\ Quirrenbach,~A.,\ Witzel,~A.,\ Krichbaum,~T.~P.,\ Hummel,~
			   C.~A.,\ \& Zensus,~J.~A.\ 1991, \aap,
			   241, 15 
\bibitem[{Racine}(1970)]{Racine70}
  Racine,~R.\ 1970, \apj, 159, L99 
\bibitem[{Raiteri} {et~al.}(2007)]{Raiteri07}
  Raiteri,~C.~M.,\ \etal\ 2007, \aap, 473, 819 
\bibitem[{Raiteri} {et~al.}(2008)]{Raiteri08}
  Raiteri,~C.~M.,\ \etal\ 2008, \aap, 491, 755 
\bibitem[{Schlegel,} {Finkbeiner} {\&} {Davis}(1998)]{Schlegel98}
  Schlegel,~D.~J.,\ Finkbeiner,~D.~P.,\ \& Davis,~M.\ 1998, \apj, 500, 525
\bibitem[{Sikora,} {Moderski} {\&} {Madejski}(2008)]{Sikora08}
  Sikora,~M.,\ Moderski,~R.,\ \& Madejski,~G.~M.\ 2008, \apj, 675,71
\bibitem[{${\rm Sillanp\ddot{a}\ddot{a}}$} {et~al.}(1993)]{Sillanpaa93}
  ${\rm Sillanp\ddot{a}\ddot{a}}$,~A.,\ Takalo,~L.~O.,\ Nilsson,~K.,\
			   \& Kikuchi,~S.\ 1993, \apss, 206, 55 
\bibitem[{Skrutskie} {et~al.}(2006)]{Skrutskie06}
  Skrutskie,~M.~F.,\ \etal\ 2006, \aj, 131, 1163
\bibitem[{Takahashi} {et~al.}(1996)]{Takahashi96}
  Takahashi,~T.,\ \etal\ 1996, \apj, 470, L89
\bibitem[{Tashiro} {et~al.}(1995)]{Tashiro95}
  Tashiro,~M.,\ Makishima,~K.,\ Ohashi,~T.,\ Inda-Koide,~M.,\
			   Yamashita,~A.,\ Mihara,~T.,\ \&
			   Kohmura,~Y. 1995, \pasj, 47, 131
\bibitem[{Urry} {\&} {Padovani}(1995)]{Urry95}
  Urry,~C.~M.,\ \& Padovani,~P.\ 1995, \pasp, 107, 803 
\bibitem[{Vercellone} {et~al.}(2008)]{Vercellone08}
  Vercellone,~S.,\ \etal\ 2008, \apj, 676, L13
\bibitem[{Vercellone} {et~al.}(2009)]{Vercellone09}
  Vercellone,~S.,\ \etal\ 2009, \apj, 690, 1018
\bibitem[{Villata} {et~al.}(2006)]{Villata06}
  Villata,~M.,\ \etal\ 2006, \aap, 453, 817
\bibitem[{Watanabe} {et~al.}(2005)]{Watanabe05}
  Watanabe,~M.,\ \etal\ 2005, \pasp, 117, 870
\bibitem[{Wolff,} {Nordsieck} {\&} {Nook}(1996)]{Wolff96}
  Wolff,~M.~J.,\ Nordsieck,~K.~H. \& Nook,M.~A. 1996, \aj, 111, 856




\end{thebibliography}
\end{document}